\documentclass[10pt,a4paper]{article}

\usepackage{amsmath}
\usepackage{amssymb}
\usepackage{hyperref}
\usepackage{cite}
\usepackage[left=2.5cm,right=2.5cm,top=2cm,bottom=2cm]{geometry}

\usepackage{tcolorbox}
\begin{document}

%\special{papersize=8.5in,11in}
%\setlength{\pdfpageheight}{\paperheight}
%\setlength{\pdfpagewidth}{\paperwidth}
%
%\conferenceinfo{SYSTOR '16}{June 6--8, 2016, Haifa, Israel}
%\copyrightyear{2016}

% Uncomment one of the following two, if you are not going for the
% traditional copyright transfer agreement.

%\exclusivelicense                % ACM gets exclusive license to publish,
                                  % you retain copyright

%\permissiontopublish             % ACM gets nonexclusive license to publish
                                  % (paid open-access papers,
                                  % short abstracts)

\title{Forman-Ricci flow for change detection \\ in large dynamic data sets}

\author{Melanie Weber$^{1}$, J{\"u}rgen Jost $^{1,2}$ and Emil Saucan$^{1,3}$\\
%% Author affiliations
\small $^{1}$Max-Planck-Institute for Mathematics in the Sciences; Inselstrasse 22, 04103 Leipzig, Germany \\
\small $^{2}$ Santa Fe Institute; 1399 Hyde Park Road Santa Fe, New Mexico 87501 USA\\
\small $^{3}$ Technion - Israel Institute of Technology; Haifa 32000, Israel\\
{\footnotesize (Contact: melweber@t-online.de, jost@mis.mpg.de, semil@ee.technion.ac.il )}
}

\maketitle

\begin{abstract}
\noindent We present a viable geometric solution for the detection of dynamic effects in complex networks. Building on Forman's discretization of the classical notion of Ricci curvature, we introduce a novel geometric method to characterize different types of real-world networks with an emphasis on peer-to-peer networks. We study the classical Ricci-flow in a network-theoretic setting and introduce an analytic tool for characterizing dynamic effects. The formalism suggests a novel computational method for change detection and the identification of fast evolving network regions and yields insights into topological properties and the structure of the underlying data.

\end{abstract}

%Categories and subject descriptors are required for the camera-ready version only. You can choose not to include this in your submission.
%\category{CR-number}{Networked, mobile, wireless, peer-to-peer, and sensor systems}{third-level}

% general terms are not compulsory anymore,
% you may leave them out
%\terms
%change detection, peer-to-peer network

%\keywords
%Ricci flow, Forman curvature, complex systems, dynamic networks

% ----------------------------
%	INTRODUCTION
% ----------------------------

\section{Introduction}
Complex networks are by now ubiquitous, both in every day life and as mathematical models for a wide range of phenomena \cite{Watts1998, Barabasi1999,  Albert2002, barabasi_net_sci} with applications in such diverse fields as Biology \cite{Barabasi2004,  Petri2014}, transportation and urban planning, \cite{Subelj2011,  Barthelemy2011, barabasi_net_sci}, social networks like Facebook and Twitter \cite{Ellison2007}, and --  in its relevance dating back to early work on networks -- in communication and computer systems \cite{Barabasi1999}. The later belong to the class of peer-to-peer networks whose structure is characterized information transfer between the ``peers''. With novel geometric methods we attempt to analyze the structural properties and dynamics of real-world networks, focusing on peer-to-peer networks as an examplary use case. 

While the main corpus of theoretical research in the analysis of networks and related structures has been focused on the properties of the various (discrete) Laplacians (see \cite{banerjee} for an overview on the state of the art), a more recent -- and, as we shall note below -- related direction concerns the geometrical characterization of real-world and model type networks, see  \cite{Ollivier2009, Jost2014, Wu2015, Sandhu2015a}. 

Given that in geometry, curvature (intuitively, a measure quantifying the deviation of a geometrical object from being flat) plays a central role, the drive for finding various discrete, expressive notions of curvature capable of describing the structure of networks is most natural \cite{Watts1998,Barabasi1999,Eckmann2002}. More recent ``geometric'' approaches consider discrete definitions of curvature as in \cite{Shavitt2004,Saucan2005,Narayan2011,Wu2015}.

While somewhat esoteric for the ``layman'', Ricci curvature plays a central role in Riemannian Geometry and has recently proven to be a remarkably powerful tool, mainly given its role in the celebrated proof by G. Perelman of Poincare's Geometerization Conjecture \cite{Perelman2002,Perelman2003}, following a path laid down earlier by R. Hamilton \cite{Hamilton}. Moreover, Ricci curvature proved itself to be unexpectedly flexible and adaptable to various discrete and more general settings, mainly in the version introduced by Y. Ollivier \cite{Ollivier2009, Ollivier2010, Ollivier2013}. Various theoretical and practical applications have been explored in the literature, see respectively \cite{Bauer2012, Jost2014, Ni2015, Sandhu2015a, Sandhu2015b}.

Despite its practicability in theoretical settings, the Ollivier-Ricci curvature is limited in its applicibility on large, real-world networks. A major constraint is the high computational cost of its calculation. The Ollivier-Ricci curvature necessitates the calculation of the so called \textit{earth mover's distance}\footnote{a.k.a. the Wasserstein  1-metric}, which in turn requires solving a non-trivial linear programming problem, thus rendering the computation unfeasible for trully large scale networks. A further limitation comes as a direct consequence of the very definition of Ollivier's concept: Given that the curvature's definition is instrinsically based on Optimal Transportation Theory, it follows that it is excellently suited for the investigation of information transfer in communication networks, yet it is less expressive as a model for structurally different types of interaction networks as occuring in biology or social sciences.

However, there is an alternative notion of discrete Ricci curvature that is computationally by far simpler and equally suited to describe the above discussed complex systems. The definition, introduced by Forman in \cite{Forman2003}, applies to a very generalized class of cellular structures that includes triangular and polygonal meshes, as well as graphs (and therefore networks). Therefore, the Forman-Ricci is a natural candidate for a discrete Ricci curvature that can be universally adapted to any type of network. After emerging as highly suited for rendering images \cite{saucan0, saucan1, saucan2} and complex network analysis \cite{Sreejit}, we will explore its applicibility on characterizing dynamic effects in complex systems.

In the present article we continue and expand the programm initiated partly by \cite{Sreejit}, namely examining the relation of Forman-Ricci curvature with other geometric network properties, such as the node degree distribution and the connectivity structure. Based on this analysis, we suggest characterization schemes that yield insights in the dynamic structure of the underlying data as described in the following section. The emphasis of the article lies on analyzing the class of peer-to-peer networks, of which we provide two examples: Email communications \cite{konect} and information exchange with the file-exchange system Gnutella \cite{leskovec,gnutella}. 

The main part of the paper introduces a novel change detection method for complex dynamic networks that exploits the Ricci flow on the edges with respect to the Forman curvature. Efficient implementations of the formalism enable structural analysis of big data as we demonstrate with the Gnutella example. For this, we walk the reader through the analysis of sets of peer-to-peer network with respect to change detection, providing an overview on the work flow of the method.

Future applications include the characterization of dynamic effects in various classes of real-world networks and the analysis of the underlying data, as well as curating of related data bases. 

% --------------------------- 
% 	METHODS
% ---------------------------

\section{Methods}
\begin{figure*}
\begin{center}
%\centerline{
\includegraphics[scale=0.25]{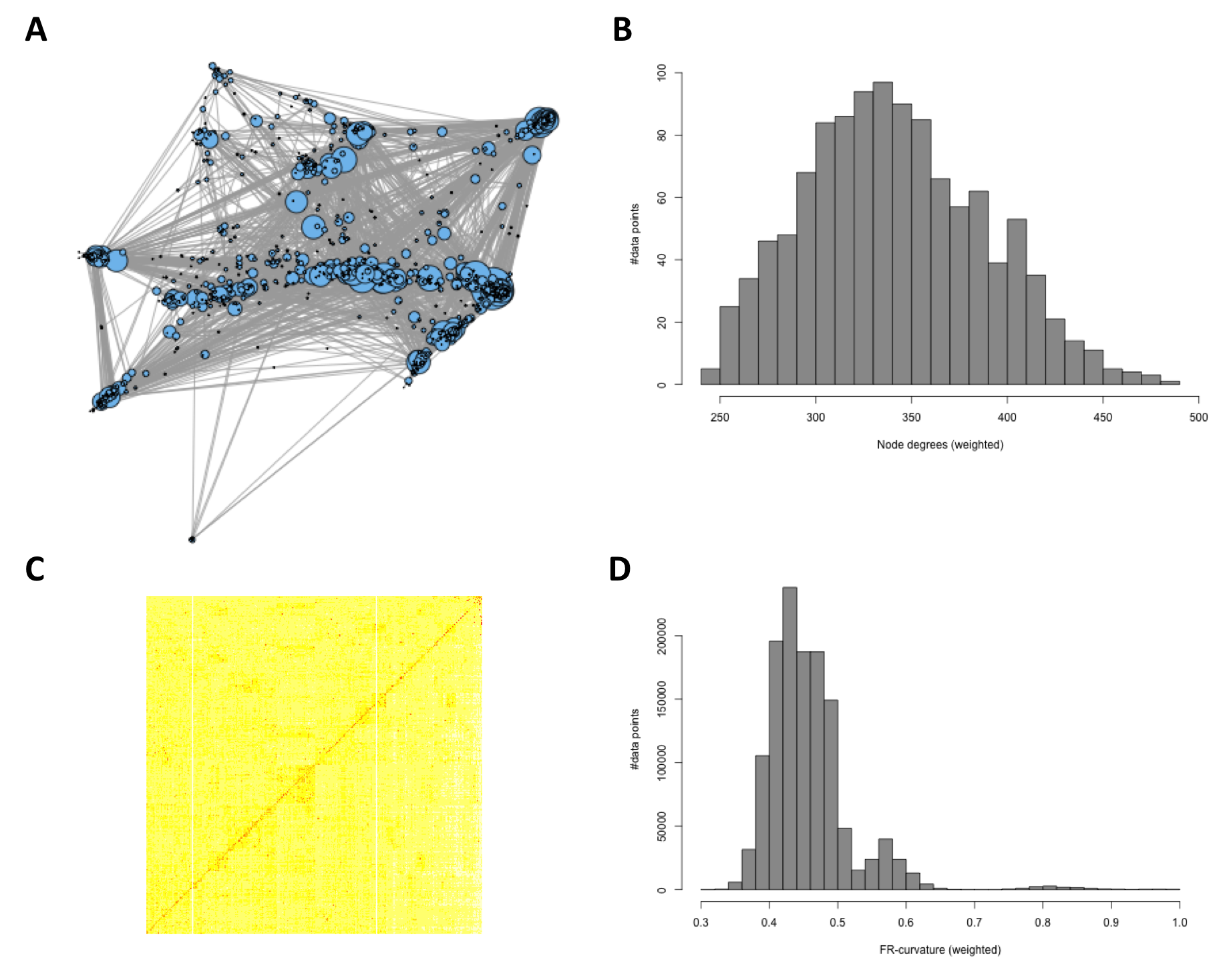} 
%}
\caption{\textsf{ Characterization of a weighted and undirected peer-to-peer network (email correspondence \cite{michalski2011,konect}) with Forman-Ricci-curvature. \textbf{A:} Network plot. Node sizes are scaled with respect to their node degrees. \textbf{B:} (Weighted) node degree distribution. \textbf{C:} Curvature map, representing the Forman-Ricci-curvature along the network's edges. \textbf{D:} Histogram showing the distribution of the Forman-Ricci-curvature.
}}
\label{email-net}
\end{center}
\end{figure*}
The change detection method introduced in this article is based on a network-analytic formulation of the Ricci-flow using a discrete Ricci-curvature on networks introduced by R. Forman \cite{Forman2003}. In this section we define both the Forman-Ricci-curvature and the Ricci-flow on networks and demonstrate their ability to characterize complex dynamic systems and their underlying data on the example of a peer-to-peer network.
\subsection{Forman-Ricci curvature on networks}
%\add[MW]{\textbf{EMIL} theory write-up, 0.5-1 page}
In the classical context of smooth Riemannian manifolds (e.g. surfaces), Ricci curvature represents an important geometric invariant that measures the deviation of the manifold from being locally Euclidean by quantifying its volume growth rate. An essential property of this curvature is that it operates directionally along vectors. For our discrete setting it follows directly that Forman's curvature is associated with the discrete analog of those vectors, namely the edges of the network.

While, as mentioned above, Forman's approach defines Ricci curvature on the very general setting of $n$-dimensional cellular structures, we will concentrate on the simpler case of 1-dimensional weighted cellular spaces that can be represented as a weighted network graph. We will not consider higher dimensional cases here, since their technicalities would carry us well beyond the scope of the present paper and our intended applications. For a theoretical introduction see \cite{Forman2003}.

In the 1-dimensional case, Forman's Ricci curvature for a network edge is defined by the following combinatorical formula

\begin{flushleft}
\begin{small}
\begin{align} \label{eq:FormanRicci1dim}
{\rm Ric}_{\bold{F}} (e) = \omega (e) \left( \frac{\omega (v_1)}{\omega (e)} +  \frac{\omega (v_2)}{\omega (e)}  - \sum_{\omega (e_{v_1}) \sim e, e_{v_2} \sim e} \left[\frac{\omega (v_1)}{\sqrt{\omega (e) \omega (e_{v_1})}} + \frac{\omega (v_2)}{\sqrt{\omega (e) \omega (e_{v_2})}} \right] \right)\,
\end{align}
\end{small}
\end{flushleft}
where
\begin{itemize}
\item $e$ denotes the edge under consideration that connects the nodes $v_1$ and $v_2$;
\item $\omega (e)$ denotes the (positive) weight on the edge $e$;
\item $\omega (v_1), \omega (v_2)$ denote the (positive) weights associated with the nodes $v_1$ and $v_2$, respectively;
\item $e_{v_1}, e_{v_2}$ denote the set of edges connected to nodes $v_1$ and $v_2$, respectively.
\label{eq:ricci-curv}
\end{itemize}

Note that in equ. (1) only edges parallel to a given edge ${\rm e}$ are taken into account, i.e. only edges that share a node with ${\rm e}$.
%\add[ES]{Should I add something about parallel edges or it would produce more confusion?}
%\add[MW]{I think a short note on this (maybe as a footnote?) would be helpful.}
We highlight that, by its very definition, Forman's discrete curvature is associated intrinsically to edges and therefore ideally suited for the study of networks. Particularly, it does not require any technical artifice in extending a node curvature measure to edges, as some other approaches do. 
%Those formalisms typically incorporate the network faces into the calculation requiring the computational search for indirect connections (i.e. pathes along edges that indirectly 'connect' not adjacent nodes) and higher order structures like triangles and quadrilaterals. The reduction of 1-dimensional components however, simplifies the calculation and makes the computational effort feasable even on very large networks.
In particular, there is no need to artificially generate and incorporate two- or higher dimensional faces: An approach that would impose severe constraints on computatbility.
 
Additionally, a ``good'' discretization of Ricci curvature, such as Forman's proves to be \cite{Forman2003}, will capture the Ricci curvature's essential characteristic of measuring the growth rate - a property that is of special interest in the context of dynamic networks. Therefore, the Forman-Ricci curvature represents a way of determining whether a networks has the potential of growing infintely (negative curvature) or can only attain a maximal - and therefore computable - size (positive curvature). 

Clearly, this aspect represents a further motivation of studying the Ricci curvature of networks, since it allows to distinguish numerically between expander type networks of essentially negative curvature, such as peer-to-peer and small world networks that are on average of strictly negative curvature (see also \cite{Ni2015}).
%\add[ES]{Am I giving away too much here or would this make things more interesting?}\add[MW]{What about adressing this with:} A follow-up article by the authors that adresses these more theoretical questions is in preparation.

%\add[MW]{This should go to the discussion section I think:}
%\textit{\begin{footnotesize}
%Another important fact, whose implications we shall discuss later one, is that, by it very definition, obtained, via an algebraic-geometric approach to the so called Bochner-Weitzenb\''{o}ck formula (see, e.g. \cite{Jost2011}), it relates between the Laplacian and Ricci curvature. More precisely, it prescribes a correction term for the standard Laplacian (or Laplace-Beltrami) operator, in terms of the curvature -- or rather curvatures -- 
%of the underlying manifold. Given that Laplacian plays a key and well known 
%role in heat equation (see, e.g. \cite{Jost2011}), it is easy to gain some basic 
%physical intuition behind phenomenon as follows: The heat evolution on
%a curved metal plate differs from that on a planar one
%in a manner that is evidently dependent on the shape
%(i.e., curvature) of the plate.
%\end{footnotesize}}

\subsection{Characterizing large data sets with Ricci curvature}
%\add[MW]{analysis of p2p email-network; results in figure panel, 1 page} \\
We now want to explore the Forman curvature as a tool for characterizing real-world networks. Since this paper centers around peer-to-peer networks, we choose an example of email communication from \cite{konect}. In such network (denoted ${\rm G}$), nodes describe correspondents and edges the exchange of messages.

To characterize the network's structure with curvature, we have to impose normalized weighting schemes

\begin{eqnarray}
\omega : V(G) \mapsto [0,1] \qquad {\rm (nodes)} \\
\gamma : E(G) \mapsto [0,1]  \qquad {\rm (edges)}
\end{eqnarray}
on both nodes and edges. Naturally, we want to weight extensively used communication channels (edges) higher than rarely used ones. For this we calculate the minimum path length ${\rm l}$ between each pair of nodes $(v_i, v_j)$ and impose

\begin{eqnarray}
\gamma (v_i, v_j) =
\begin{cases}
\frac{1}{l}, & l \leq 6, v_i \sim v_j \\
0,  & {\rm else.}
\end{cases}
\end{eqnarray}
The motivation behind this choice of weights, lies in the ``small world''-property (i.e. a maximum degree of separation of six\cite{DESOLAPOOL19785}) that has been reportedly found in real-world. Accounting for this, we only check for indirect connections up to a path length of six and scale the weights according to the distribution.

Analogousely, we impose a weighting scheme on the nodes. Naturally, the busiest communicators should have the highest weights. Therefore we choose a combinatorial weighting scheme based on node degree, i.e. the number and weight of connections for each node $v$:

\begin{eqnarray}
\omega (v) = \frac{1}{{\rm deg}(v)} \sum_{e_i \sim v} {\gamma (e_i)}\,.
\end{eqnarray}
The resulting structure of the network is shown in Fig. \ref{email-net}.A, the size of the nodes reflects their weights.

Using edge and node weights, we can now determine the Forman curvature distribution in the network. A curvature map (Fig. \ref{email-net}.C) provides a planar representation that highlights clusters and distinguished regions. The histogram (Fig. \ref{email-net}D) shows the distribution of the curvature values allowing for comparison with the node degree distribution (Fig. \ref{email-net}.B). With indicating a correlation between the two distributions, the results highlight the strong influence of node degree weights on the network's topology. This is consistent with observations in email communications: Densly interconnected communities form around busy communication channels and active correspondents. 

\subsection{Ricci-flow with Forman curvature}
%\add[MW]{\textbf{EMIL} theory write-up: definition of Ricci flow, 0.5-1 page}\\
As mentioned earlier, the Ricci flow as a powerful geometric tool was devised by Hamilton\cite{Hamilton} and further developed by Perleman\cite{Perelman2002, Perelman2003} in the course of his celebrated proof of the Poincare conjecture. Since then, it has continued to be an active and productive field of study, both in terms of theoretical questions, but also for diverse practical applications, including work by Gu et al. (see, e.g. \cite{Sarkar2009}). Those mainly build on a combinatorial version introduced by Chow and Luo \cite{Chow2003}. However, other discretizations of the flow, with reported applications in network and imaging sciences, are explored in the literature \cite{Ni2015,Saucan2014a}. 

The classical Ricci flow is defined by

\begin{equation}
{\frac{\partial g_{ij}}{\partial t} = - {\rm Ric}_{\bold{F}} (g_{ij}) \cdot g_{ij}}
\end{equation}
where $g_{ij}$ denotes the metric of the underlying manifold, represented by the earlier introduced weighting scheme of the network's edges.

The reader might note the resemblance with the classical Laplace (or more precisely Laplace-Beltrami) flow that has become, by now, standard in Imaging and Graphics (see, e.g. \cite{DESOLAPOOL19785, Xu04} and the references therein) being defined as

\begin{equation}
{\frac{\partial I}{\partial t} = \Delta I}
\end{equation}
where $I$ denotes the image, viewed as a parametrized surface in $\mathbb{R}^3$ (and $\Delta$ denotes, as customary, the Laplacian)\footnote{The difference in sign is due to the two different conventions common when defining the Laplacian}.

The resemblance is neither accidental nor superficial. Indeed, the Ricci curvature can be viewed as a Laplacian of the metric. We shall address the practical implications of this observation in the sequel and a follow-up article by the authors that adresses more theortical questions of this matter.

%\add[ES]{Should I really formally define it? - Perhaps such details better left for the journal (follow-up) paper - here we might just confuse the reader (and we are not actually treating this flow here).}

In our discrete setting, lengths are replaced by the (positive) 
%\add[ES]{Did we (me) specify the weights are positive?!...}\add[MW]{yes, see previous section} 
edge weights. Time is assumed to evolve in discrete steps and each "clock" has a length of 1. With these constraints the Ricci flow takes the form

\begin{equation}
\tilde{\gamma} (e)  - \gamma (e) = - {\rm Ric}_{\bold{F}} (\gamma (e)) \cdot \gamma (e)\,,%
\end{equation}
where $\tilde{\gamma} (e)$ denotes the new (updated) value of $\gamma (e)$ with $\gamma (e)$ being the original - i.e. given - one.

%\add[ES]{I think it is best to use this form for the Forman-Ricci curvature, not to produce confusion, with the general one.}

In this context, we want to point out and discuss a few issues and observations regarding this last equation:\newline
%\add[ES]{You might keep all -- or just some/some part -- of the comments below (perhaps not itemized, though.}

\begin{enumerate}
\item At each iteration step (i.e. in the process of updating $\tilde{\gamma}(e)$ to $\gamma (e)$), the Forman curvature has to be recomputed for each edge $e$, since it depends on its respective weight $\gamma (e)$. This clearly increases the computational effort on magnitudes, however, the computation task is less formidable than it might appear at first.

\item As already stressed, we consider a discrete time model. Since for smoothing (denoising) a short time flow is to be applied\footnote{Because, by the theory for the smooth case, a long time flow will produce a limiting state of the network.} only a small number of iterations need be considered. The precise number of necessary iteration is to be determined experimentally. Even though a typical number can be found easily, best results may be obtained for slightly different numbers -- depending on the network, and the type and level of the noise, of course.
     
\item Forman also devised a continuous flow \cite{Forman2003}. In the context of the present article, a continous setting is not required, but for other types of networks, where the evolution is continuous in time, it might be better to implement the other variant.
%
%\add[ES]{Actually, I think is dangerous to mention it here - better to leave it for the later work.}
\end{enumerate}
In addition to the Ricci flow above, one can consider the {\em scalar curvature flow} that in our case will have the form:
\begin{equation}
\frac{d}{dt} \gamma (e_i) = - \frac{1}{2}\big( {\rm scal}_F(v) - {\rm scal}_F(v_i)\big) \gamma (e_i)\,,%
\end{equation}
where $e_i = e_i(v, v_i)$ denote the edges through the node $v$, and ${\rm scal}_F(v)$ the (Forman) scalar curvature at a node $v$, which we define by

\begin{equation}
{\rm scal}_F(v) = \sum_{e_i \sim v} {\rm Ric}_F (\gamma (e_i)) \,.
\end{equation}
%
%\add[ES]{Did we do this (anywhere)?... Should I mention it?...}

\subsection{Characterizing dynamic data with Ricci Flow}
%\add[MW]{describe dynamic analysis, transition to next chapter}\\
The introduced Ricci-flow can be utilized to characterize dynamic data. Given snapshots of a system at various (discrete) times $\lbrace t_i \rbrace$, we analyze the Ricci flow on the corresponding network representations. The flow yields insights into structural changes providing a tool to identify "interesting" network regions. Applications include efficient change detection in large dynamic data representing complex systems, as described in the following section.\\
Let $\left( t_i \right)_{i \in I}$ be a discrete time series with step size $\Delta t$. Consider a complex dynamic system of whose behavior we have snapshots at times $(t_i)$ represented in weighted network graphs $(G_i)_{i \in I}$. We disregard any self-loops, i.e. edges connecting a node with itself. Let $t_n$ and $t_{n+1}$ be consecutive time points with corresponding graphs $G_n$ and $G_{n+1}$. The weighting schemes of the nodes

\begin{eqnarray}
\omega_{n,n+1}^0: V(G) \mapsto [0,1]
\end{eqnarray}
and edges

\begin{eqnarray}
\gamma_{n,n+1}^0: E(G) \mapsto [0,1] 
\end{eqnarray}
characterize the topological structure of the system at times $t_n$ and $t_{n+1}$. We can estimate the Ricci-flow for the time step $\Delta t$ by iterating $K$ times over

\begin{small}
\begin{eqnarray}
\gamma_{n,n+1}^{k+1} (e_i) = \gamma_{n,n+1}^{k} (e_i) - \Delta t \cdot {\rm Ric}_F (\gamma_{n,n+1}^{k} (e_i) ) \cdot \gamma_{n,n+1}^{k} (e_i)
\label{eq:Ricci-flow}
\end{eqnarray}
\end{small}
resulting in modified weighting schemes $\gamma_n^K$ and $\gamma_{n+1}^{K}$ respectively. The correlation between these weighting schemes characterizes the flow. Regions that were subject to significant change during the time $\Delta t$ can be identified by thresholding the resulting correlation matrix.

%Following method for detecting changes in images \citep{saucan0} using curvature, detect changes in FR-map for large data sets:
%\begin{enumerate}
%\item Calculate FR-curvature heatmap for network state at times $t$ and $t+\Delta t$ (e.g. version of a data base, downloaded at $t_1$ and its preceeding version from $t_2 = t_1 + \Delta t$ ).
%\item Normalize the pair of networks.
%\item Calculate the discrete Ricci-flow for each network through iteration over 
%\[g_{ij}^{k+1} = g_{ij}^{k} - \Delta t \cdot Ric(g_{ij}^{k})\]
%with FR-curvature $Ric$ ($g^0$ refers to the original weighting scheme of the network).
%
%\add[ES]{ I think that here we should either explain that this is a short-time flow (but perhaps not say precisely how we compute...) or use the correct formula - and apply it (but for this is no time...) I'll add this myself once we decide.} \\
%\add[MW]{This is just an outline, I only worked a bit on the write-up of the introduction section. Formulas, short-time arguments etc. will definetely be included here.}
%
%\item Determine correlation between the resulting Ricci metrices of both network states. Changes can be detected when applying an adaptive threshold.
%\item Track unusual changes/ rare events in dynamic network states in discrete time.
%\end{enumerate}

% -----------------------
% Applications
% -----------------------

\section{Application}
\begin{figure*}
\begin{center}
\centerline{
\includegraphics[scale=0.2]{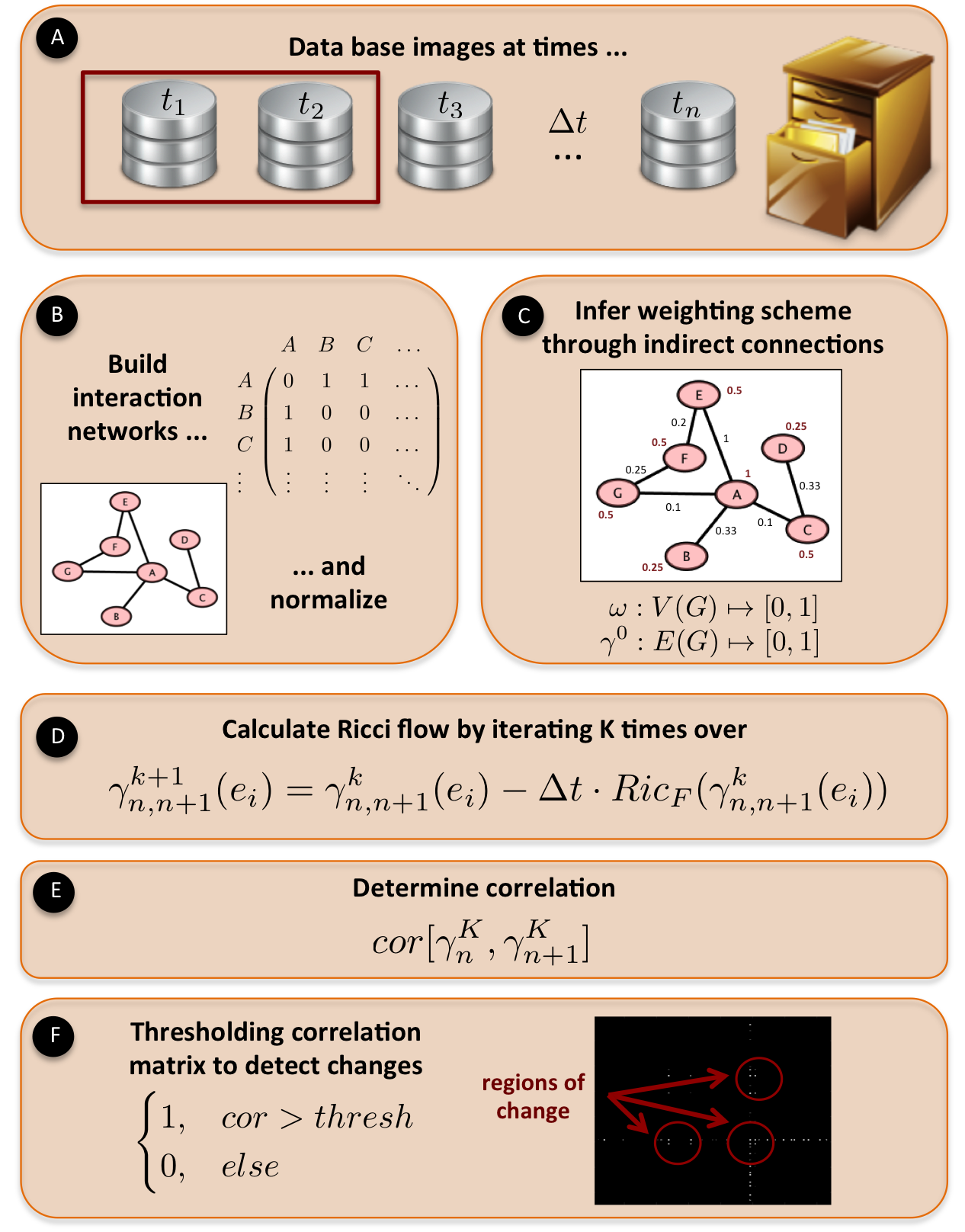} 
}
\caption{\textsf{ {\footnotesize 
Schematic overview of the change detection workflow. \textbf{A:} We consider a set of "snapshots" of a system or data base at discrete times $(t_i)_{i \in I}$ with step size $\Delta t$. \textbf{B:} We infere unweighted (binary) networks from each snapshot that represent the structure of the underlying data. To simplify comparison, we normalize all networks. \textbf{C:} By superimposing weighting schemes based on "indirect connections", we extend the unweighted networks to weighted ones (see section 2). \textbf{D:} We calculate the Ricci flow for each time step $\Delta t$ by iterating over a the given scheme. \textbf{E:} For detection of changes, we compare the final (smoothed) weighting schemes after $K$ iterations. \textbf{F:} To identify regions that were subject to significant change, we threshold the corrleation matrix obtained in (E). Light regions in the resulting map indicate such regions.
}}}
\label{scheme}
\end{center}
\end{figure*}
%
%%%%%%%%%%%%%%%%%
%
\begin{figure*}
\begin{center}
%\centerline{
\includegraphics[scale=0.25]{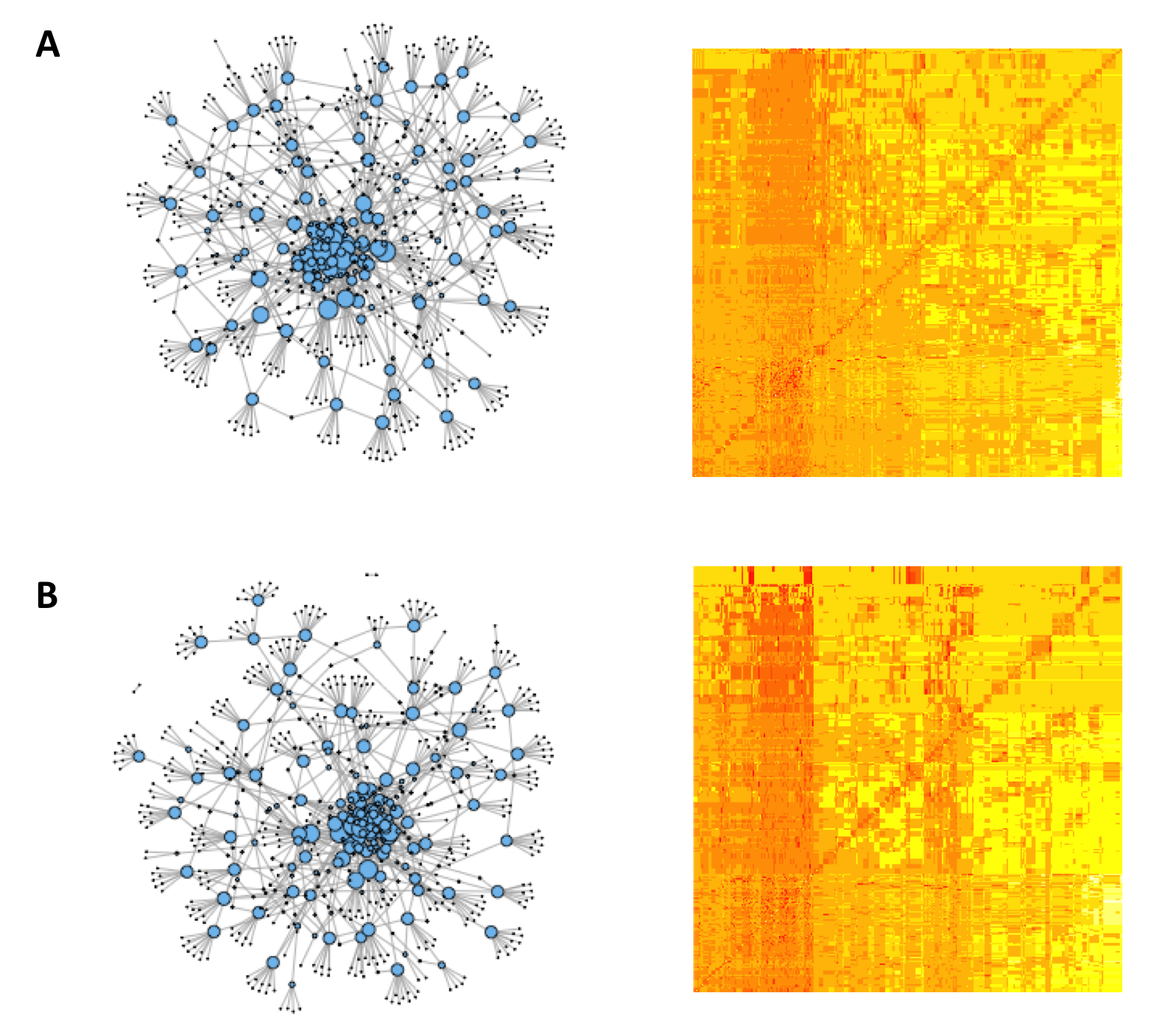} 
%}
\caption{We analyze an Internet peer-to-peer network representing Gnutella file exchanges on two consecutive days (August 8 and August 9 2002, from \cite{leskovec,gnutella}). \textbf{A:} Network plot (left) and curvature map (right) displaying the distribution of Forman curvature for a Gnutella snapshot at August 8. \textbf{B:} Analogous plots for a Gnuttela snapshot from the following day, August 9.
}
\label{gnutella}
\end{center}
\end{figure*}
%
%%%%%%%%%%%%%%%%%%%
\subsection{Change detection with Ricci flow}
A major challenge of modern data science lies in characterizing dynamic effects in large data sets, such as structural changes in discrete time series of "snapshots" of the state a system or frequently updated large data bases. Commonly, network graphs are infered from data representing interactions and associations in the underlying data. In the case of peer-to-peer networks, the network describes the information flow (edges) between the peers (nodes). 

We want to use the Forman-Ricci curvature to analyze dynamic changes in the structure of such interaction networks infered from large data sets. Specifically, we want to characterize the information flow between the network's nodes using the previousely introduced Ricci flow. Our method follows formalism described in section 2 and is schematically displayed in Fig. \ref{scheme}.

The analysis of the information flow can be used to detect changes or distinguished regions of activity in the data. Applications include the curation of large open-access data bases and the detection of rare events in experimental data, such as spiking neurons in measurements of neuronal activity.

%\begin{itemize}
%\item \textbf{Brief summary of the problem: Efficient changes in networks representing large data sets/ complex systems (e.g. in open-access data bases, hinting on biased underlying data/ studies.}

%\add[ES]{ Are we actually doing denoising here? - I think not. Therefore, better to leave this for the Future Work section.}
%\add[MW]{When brainstorming two weeks ago, we wanted to include the denoising, that's why its part of the outline. But I agree, we will probably not get to the point to include this until friday.}

%\item \textbf{Change detection} in data base version (methods \& implementation).
%\item \textbf{Use case:} Detect changes in gnutella (\cite{leskovec,gnutella}, versions Aug 8 and Aug 9 2015).
%\end{itemize}

\subsection{Analysis of Gnutella peer-to-peer network}
\begin{figure*}
\begin{center}
%\centerline{
\includegraphics[scale=0.15]{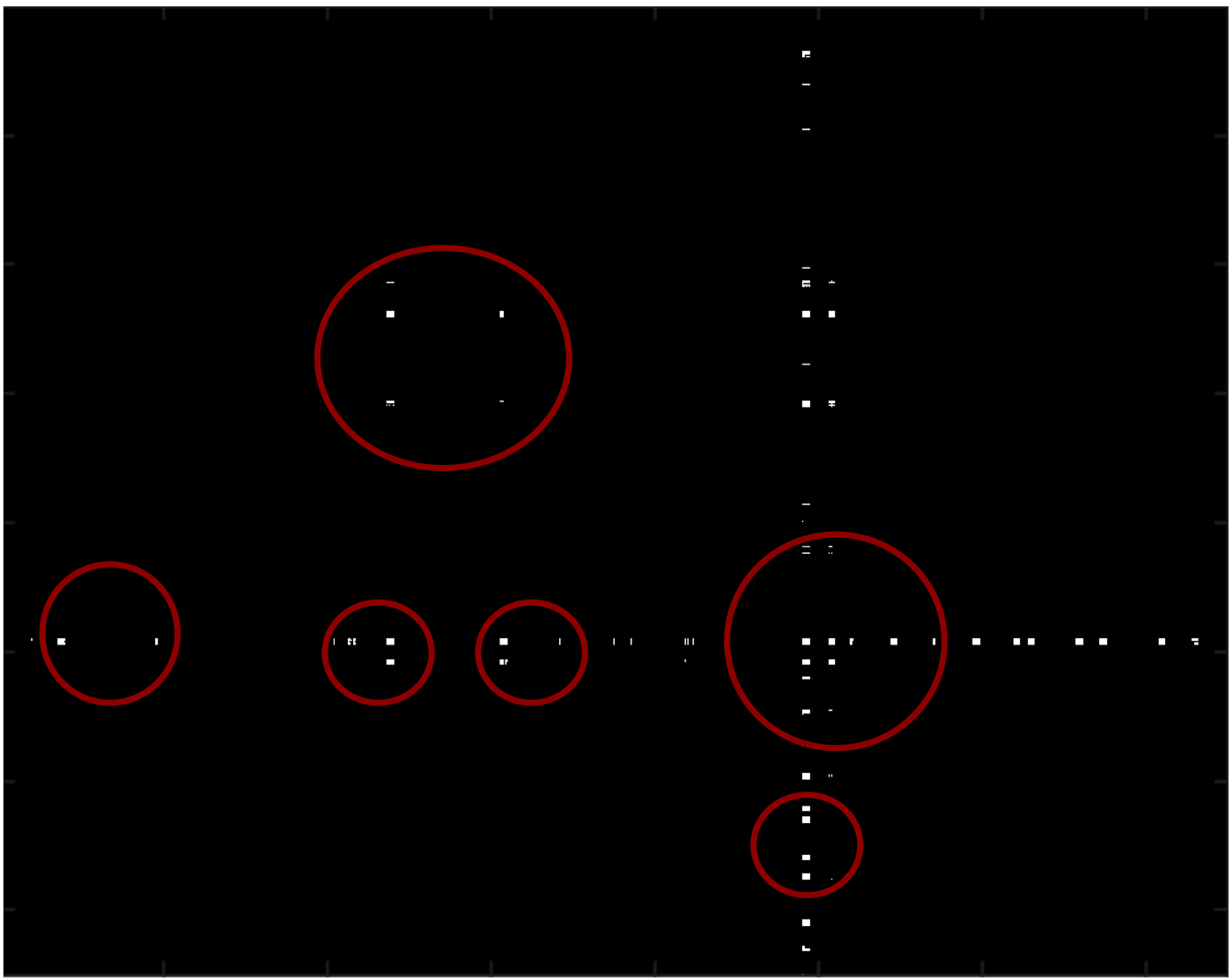} 
%}
\caption{\textsf{Detection of changes and regions of distinguished activity in Gnutella snapshots of two consecutive days \cite{leskovec,gnutella} with Ricci flow and a parameter choice of $K=10$ and $\Delta t= 0.8$. The highlighted spots correspond to single edges and assortments of edges and nodes (clusters) with high activity (flow).
}}
\label{change-detect}
\end{center}
\end{figure*}

We consider a series of discrete time snapshots of a complex peer-to-peer system, represented as network graphs $(G_i)_{i \in I}$. To characterize the Ricci flow, we apply the earlier described formalism pairwise, i.e. we iterate over (\ref{eq:Ricci-flow}) for snapshots $(G_n, G_{n+1})$ at consecutive time points $t_n$ and $t_{n+1}$. Here we use the file-exachange service \textit{Gnutella} as an example, analyzing the peer-to-peer networks resulting from exchange activities on two consecutive days (August 8 and August 9 2002, \cite{leskovec,gnutella}). Fig. \ref{gnutella} shows the networks infered from the data sets and the corresponding curvature maps with the distribution of Forman Ricci curvature. We applied our change detection method with a correlation threshold of $t=0.9$ to detect regions of significant changes, i.e. groups of peers with significant activity. The results are shown in Fig. \ref{change-detect}, represented as a heatmap. 
Light spots in the heatmap correspond to network regions with large flow and structural changes allowing for detection and localization of such dynamic effects. The method provides a clear representation of dynamic changes in networks, especially in terms of the community structure of the network. Given the intrinsic community structure of a network (clusters), which is also hinted in the Forman curvature maps, one can examine the influence of the flow on this specific structural property..

% ---------------------
% DISCUSSION
% ---------------------

\section{Discussion and Future Work}

The dawning age of big data opens up great numbers of possibilities and perspectives to gain insights in the principles of nature, humanity and technology through collection and (statistical) analyis of data. However, the vaste amount of available data challenges our analysis methods leading to an increased need of automated tools that can perform rapid and efficient data evaluations. 

Networks are an efficient and commonly used data representation, emphasizing on interactions and associations. This representation is ideal for analyzing structures within the data under geometric aspects. We have used networks to characterize peer-to-peer systems and analyze dynamic effects in information transfer between its peers. Particularly, we introduced a method for detecting changes in these dynamics. Possible applications of this theoretical framework include denoising of experimental data and identification of "interesting" groups of data points and activities of network regions. 

The geometric methods we used in the present article are build on R. Forman's work on Ricci curvature in networks and the corresponding Ricci flow based on this definition. Future work, especially a follow-up article by the authors, will expand the range of geometric tools used in the methods and develop a deeper understanding of theoretical aspects. In what follows, we want to name and discuss some extensions and theoretical considerations that we shall adress in future work.

%\begin{enumerate}
%\item \textbf{Computational challenges in analyzing large data sets} with the formalism.
%\item Describing \textbf{challenges in (meta-) analysis of} highly noisy \textbf{biological data}. Reference former approaches and relevant work (especially Gillis lab).
%\item impact of indirect connection in data analysis: curvature only on edges - minimizes biasing impact when analyzing infered interaction networks; for a closer look on this problem in a biological context see (Gillis, Pavlidis 2011)
%\item \textbf{Future directions}.\\
%\end{enumerate}
%
%\add[MW]{sketch: Emil}\\
One important fact, whose implications we shall discuss later on, is the so called 
Bochner-Weitzenboeck formula (see, e.g. \cite{Jost2011}), which relates graph Laplacian and Ricci curvature through an algebraic-geometric approach. More precisely, it prescribes a correction term for the standard Laplacian (or Laplace-Beltrami) operator, in terms of the curvature 
of the underlying manifold. Given that the Laplacian plays a key role in the heat equation (see, e.g. \cite{Jost2011}), it is easy to gain some basic 
physical intuition behind the phenomenon of curvature and flow: The heat evolution on a curved metal plate differs from that on a planar one in a manner that is evidently dependent on the shape (i.e., curvature) of the plate.

This conducts us to a number of prospective future directions:

\begin{enumerate}

\item A task that is almost self evident, is to further experiment with very large data sets (numbers of data points in the order of ten thousand and more);

\item Another natural target is the use of our method on different types of networks, with special emphasis on Biological Networks;

\item A statistical analysis regarding the Ricci flow, similar to the one presented here and in \cite{Sreejit}, should also be performed on various standard types of networks in order to confirm and calibrate the characterization and classifying capabilities of the Ricci curvature and flow.

\end{enumerate}
Slightly more demanding are future experiments and comparisons with the related flows, namely

\begin{enumerate}

 \item The Forman curvature versions of the scalar and Laplace-Beltrami flows. Especially the last one seems to be quite promising for network denoising, as applications  of the analogous flow in image processing showed \cite{Saucan2008}. Moreover, the Forman-Ricci curvature comes naturally coupled with a fitting version of the so called {\em Bochner Laplacian} (and yet with another, intrinsically connected, {\em rough Laplacian}). This aspect is subject to currently ongoing work and will be covered in a forthcoming paper by the authors.
 
\item Here, as for the short time Ricci flow, statistical analysis should be undertaken to validate the classification potential of the long term Ricci flow. A more ambitious, yet still feasible, future direction would be to explore network stability by considering the {\em long time} Forman-Ricci flow (as opposed to the short time one employed for denoising). This approach would exploit, in analogy with the smooth case \cite{Perelman2002,Perelman2003}, the propensity of the Ricci flow to preserve and quantify the overall, global Geometry (i.e. curvature) and essential topology of the network. This would allow us to study the evolution of a network ``under its own pressure'' and to detect and examine catastrophic events as virus attacks and denial of service attempts. Given the basic numerical simplicity of our method, this approach might prove to be an effective alternative to the Persistent Homology method (see, e.g. \cite{Petri2014}) for the 1-dimensional case of networks. Moreover, the Ricci flow does not need to make appeal to higher dimensional structures (namely {\em simplicial complexes}) that are necessary for the Persistent Homology based applications, with clear computational -- but also theoretically rigor -- advantages. The potential benefits of the Ricci flow would be, in fact, even greater, given that it deals with weighted networks, not just with their unweighted, combinatorial counterparts used in Persistent Homology.

\end{enumerate}

\section*{Acknowledgements}
ES acknowledges the warm support of the Max-Planck-Institute for Mathematics in the Sciences, Leipzig, for the warm hospitality during the writing of this paper. He would also like to thank Areejit Samal for many interesting discussions. MW was supported by a scholarship of the Konrad Adenauer Foundation.

\bibliographystyle{plain}

\bibliography{paper}

\begin{thebibliography}{10}

\bibitem{Albert2002}
R.~Albert and A.-L. Barab{\'a}si.
\newblock Statistical mechanics of complex networks.
\newblock {\em Reviews of Modern Physics}, 74(1):47, 2002.

\bibitem{banerjee}
A.~Banerjee and J.~Jost.
\newblock Spectral plot properties: Towards a qualitative classification of
  networks.
\newblock In {\em In European Conference on Complex Systems}, 2007.

\bibitem{barabasi_net_sci}
A.-L. Barabasi.
\newblock {\em Network Science}.
\newblock Cambridge University Press, 2016.

\bibitem{Barabasi1999}
A.-L. Barab{\'a}si and R.~Albert.
\newblock Emergence of scaling in random networks.
\newblock {\em Science}, 286(5439):509--512, 1999.

\bibitem{Barabasi2004}
A.~L. Barab\'{a}si and Z.~N. Oltvai.
\newblock Network biology: understanding the cell's functional organization.
\newblock {\em Nature Reviews Genetics}, 5(2):101--113, 2004.

\bibitem{Barthelemy2011}
M.~Barth{\'e}lemy.
\newblock Spatial networks.
\newblock {\em Physics Reports}, 499(1):1--101, 2011.

\bibitem{Bauer2012}
F.~Bauer, J.~Jost, and S.~Liu.
\newblock Ollivier-ricci curvature and the spectrum of the normalized graph
  laplace operator.
\newblock {\em Mathematical research letters}, 19:1185--1205, 2012.

\bibitem{Chow2003}
B.~Chow and F.~Luo.
\newblock Combinatorial ricci flows on surfaces.
\newblock {\em Journal of Differential Geometry}, 63(1):97--129, 2003.

\bibitem{saucan1}
E.~Saucan E.~Appleboim and Y.~Y. Zeevi.
\newblock Ricci curvature and flow for image denoising and superesolution.
\newblock In {\em Proceedings of EUSIPCO 2012}, volume~., pages 2743--2747,
  2012.

\bibitem{saucan0}
G.~Wolansky E.~Saucan, E.~Appleboim and Y.~Y. Zeevi.
\newblock Combinatorial ricci curvature and laplacians for image processing.
\newblock In {\em Proceedings of CISP'09}, volume~2, pages 992--997, 2009.

\bibitem{saucan2}
E.~Appelboim E.~Sonn, E.~Saucan and Y.~Y. Zeevi.
\newblock Ricci flow for image processing.
\newblock In {\em IEEEI 2014}, volume~., page~., 2014.

\bibitem{Eckmann2002}
J.~Eckmann and E.~Moses.
\newblock Curvature of co-links uncovers hidden thematic layers in the world
  wide web.
\newblock {\em Proceedings of the National Academy of Sciences},
  99(9):5825--5829, 2002.

\bibitem{Ellison2007}
N.~B. Ellison, C.~Steinfield, and C.~Lampe.
\newblock The benefits of facebook friends: Social capital and college
  students’ use of online social network sites.
\newblock {\em Journal of Computer-Mediated Communication}, 12(4):1143--1168,
  2007.

\bibitem{Sreejit}
R.~P.~Sreejith et~al.
\newblock Forman curvature for complex networks.
\newblock {\em arXiv:1603.00386}, 2016.

\bibitem{Sarkar2009}
R.~Sarkar et~al.
\newblock Greedy routing with guaranteed delivery using ricci flows.
\newblock In {\em Proc. of the 8th International Symposium on Information
  Processing in Sensor Networks (IPSN'09)}, volume~., pages 121--132, 2009.

\bibitem{forman}
R.~Forman.
\newblock Bochner's method for cell complexes and combinatorial ricci
  curvature.
\newblock {\em Discrete and Computational Geometry}, 29(3):323--374, 2003.

\bibitem{Forman2003}
R.~Forman.
\newblock Bochner's method for cell complexes and combinatorial ricci
  curvature.
\newblock {\em Discrete and Computational Geometry}, 29(3):323--374, 2003.

\bibitem{Hamilton}
R.~S. Hamilton.
\newblock The ricci flow on surfaces. mathematics and general relativity.
\newblock In {\em Contemp. Math, Amer. Math. Soc.}, volume~71, pages 237--262,
  1986.

\bibitem{leskovec}
J.~Kleinberg J.~Leskovec and C.~Faloutsos.
\newblock {Graph Evolution: Densification and Shrinking Diameters}.
\newblock In {\em ACM Transactions on Knowledge Discovery from Data (ACM
  TKDD)}, volume 1(1), 2007.

\bibitem{Jost2011}
J.~Jost.
\newblock {\em Riemannian Geometry and Geometric Analysis}.
\newblock Springer, 2011.

\bibitem{Jost2014}
S.~Jost, J.and~Liu.
\newblock Ollivierís ricci curvature, local clustering and curvature-dimension
  inequalities on graphs.
\newblock {\em Discrete \& Computational Geometry}, 51(2):300--322, 2014.

\bibitem{konect}
J.~Kunegis.
\newblock Konect - the koblenz network collection.
\newblock In {\em Proc. Int. Conf. on World Wide Web Companion}, pages
  1343--1350, 2013.

\bibitem{DESOLAPOOL19785}
l.~de Sola~Pool and M.~Kochen.
\newblock Contacts and influence.
\newblock {\em Social Networks}, 1(1):5 -- 51, 1978.

\bibitem{gnutella}
I.~Foster M.~Ripeanu and A.~Iamnitchi.
\newblock {Mapping the Gnutella Network: Properties of Large-Scale Peer-to-Peer
  Systems and Implications for System Design}.
\newblock In {\em IEEE Internet Computing Journal}, 2002.

\bibitem{michalski2011}
R.~Michalski and S.~P. Palus.
\newblock Matching organizational structure and social network extracted from
  email communication.
\newblock In {\em Lecture Notes in Business Information Processing}, volume~87,
  pages 197--206. Springer Berlin Heidelberg, 2011.

\bibitem{Narayan2011}
O.~Narayan and I.~Saniee.
\newblock Large-scale curvature of networks.
\newblock {\em Physical Review E}, 84(6):066108, 2011.

\bibitem{Ni2015}
C.~Ni, Y.~Lin, J.~Gao, X.~D. Gu, and E.~Saucan.
\newblock Ricci curvature of the internet topology.
\newblock In {\em IEEE Conference onComputer Communications (INFOCOM)}, pages
  2758--2766. IEEE, 2015.

\bibitem{Ollivier2009}
Y.~Ollivier.
\newblock Ricci curvature of markov chains on metric spaces.
\newblock {\em Journal of Functional Analysis}, 256(3):810--864, 2009.

\bibitem{Ollivier2010}
Y.~Ollivier.
\newblock A survey of ricci curvature for metric spaces and markov chains.
\newblock {\em Probabilistic approach to geometry}, 57:343--381, 2010.

\bibitem{Ollivier2013}
Y.~Ollivier.
\newblock A visual introduction to riemannian curvatures and some discrete
  generalizations.
\newblock {\em Analysis and Geometry of Metric Measure Spaces: Lecture Notes of
  the 50th S{\'e}minaire de Math{\'e}matiques Sup{\'e}rieures (SMS),
  Montr{\'e}al, 2011}, pages 197--219, 2013.

\bibitem{Perelman2002}
G.~J. Perelman.
\newblock The entropy formula for the ricci flow and its geometric
  applications.
\newblock {\em arXiv:math/0211159}, 2002.

\bibitem{Perelman2003}
G.~J. Perelman.
\newblock Ricci flow with surgery on three-manifolds.
\newblock {\em arXiv:math/0303109}, 2003.

\bibitem{Petri2014}
G.~Petri, P.~Expert, F.~F.~Turkheimer, R.~Carhart-Harris, D.~Nutt, P.~J.
  Hellyer, and F.~Vaccarino.
\newblock Homological scaffolds of brain functional networks.
\newblock {\em Journal of The Royal Society Interface}, 11(101):20140873, 2014.

\bibitem{Sandhu2015a}
R.~Sandhu, T.~Georgiou, E.~Reznik, L.~Zhu, I.~Kolesov, Y.~Senbabaoglu, and
  A.~Tannenbaum.
\newblock Graph curvature for differentiating cancer networks.
\newblock {\em Scientific Reports}, 5, 2015.

\bibitem{Sandhu2015b}
R.~Sandhu, T.~Georgiou, and A.~Tannenbaum.
\newblock Market fragility, systemic risk, and ricci curvature.
\newblock {\em arXiv:1505.05182}, .

\bibitem{Saucan2014a}
E.~Saucan.
\newblock A metric ricci flow for surfaces and its applications.
\newblock {\em Geometry, Imaging and Computing}, 1(2):259--301, 2014.

\bibitem{Saucan2005}
E.~Saucan and E.~Appleboim.
\newblock Curvature based clustering for dna microarray data analysis.
\newblock In {\em Pattern Recognition and Image Analysis}, pages 405--412.
  Springer, 2005.

\bibitem{Saucan2008}
E.~Saucan, E.~Appleboim, G.~Wolanski, and Y.~Zeevi.
\newblock Combinatorial ricci curvature for image processing.
\newblock In {\em MICCAI 2008 Workshop Manifolds in Medical Imaging: Metrics,
  Learning and Beyond}, 2008.

\bibitem{Shavitt2004}
Y.~Shavitt and T.~Tankel.
\newblock On the curvature of the internet and its usage for overlay
  construction and distance estimation.
\newblock In {\em INFOCOM 2004. Twenty-third AnnualJoint Conference of the IEEE
  Computer and Communications Societies}, volume~1. IEEE, 2004.

\bibitem{Subelj2011}
L.~{\v{S}}ubelj and M.~Bajec.
\newblock Robust network community detection using balanced propagation.
\newblock {\em The European Physical Journal B}, 81(3):353--362, 2011.

\bibitem{Watts1998}
D.~J. Watts and S.~H. Strogatz.
\newblock Collective dynamics of small-world networks.
\newblock {\em Nature}, 393(6684):440--442, 1998.

\bibitem{EGAD}
M.~Weber, S.~Ballouz, P.~Pavlidis, and J.~Gillis.
\newblock Egad: Ultra-fast functional analysis of gene networks.
\newblock {\em preprint}, 2016.

\bibitem{Wu2015}
Z.~Wu, G.~Menichetti, C.~Rahmede, and G.~Bianconi.
\newblock Emergent complex network geometry.
\newblock {\em Scientific Reports}, 5, 2015.

\bibitem{Xu04}
G.~Xu.
\newblock Discrete laplace–beltrami operators and their convergence.
\newblock {\em Computer Aided Geometric Design}, 21:767–--784, 2004.

\end{thebibliography}

\end{document}